\documentclass{evn2004}
\usepackage{txfonts}
\usepackage{graphicx}
\usepackage{natbib}
\begin{document}
   \title{Interstellar scintillation as a probe of microarcsecond scale  
   structure in quasars}

   \author{H.E. Bignall\inst{1}
          \and
          D.L. Jauncey\inst{2}
          \and
          J.E.J. Lovell\inst{2}
          \and
          L. Kedziora-Chudczer\inst{3}
          \and
          J-P. Macquart\inst{4}
          \and 
          A.K. Tzioumis\inst{2}  
          \and
          B.J. Rickett\inst{5}
          \and
          R. Ojha\inst{2}
          \and 
          S. Carter\inst{6}
          \and
          G. Cim\'o\inst{6}
          \and
          S. Ellingsen\inst{6}
          \and
          P.M. McCulloch\inst{6}
          }

   \institute{Joint Institute for VLBI in Europe, Postbus 2, 7990 AA
         Dwingeloo, The Netherlands
         \and
             Australia Telescope National Facility, PO Box 76, Epping, NSW
1710, Australia
         \and
             Institute of Astronomy, School of Physics A28, The
             University of Sydney, NSW 2006, Australia
         \and
             Kapteyn Astronomical Institute, University of Groningen,
             Postbus 800, 9700 AV Groningen, The Netherlands
         \and
             Department of Electrical and Computer Engineering,
         University of California at San Diego, La Jolla, CA 92093, USA
         \and
             School of Mathematics and Physics, University of Tasmania,
         Private Bag 37, Hobart, Tas 7001, Australia
             }

   \abstract{ Observations over the last two decades have shown that a
significant fraction of all flat-spectrum, extragalactic radio sources
exhibit flux density variations on timescales of a day or less at
frequencies of several GHz. It has been demonstrated that interstellar
scintillation (ISS) is the principal cause of such rapid
variability. Observations of ISS can be used to probe very compact,
microarcsecond-scale structure in quasar inner jets, as well as properties
of turbulence in the local Galactic ISM. A few sources show unusually
rapid, intra-hour variations, evidently due to scattering in very
nearby, localized turbulence. We present recent findings for the rapidly
scintillating quasar PKS~1257$-$326. The large-scale MASIV VLA Survey
showed that such sources are extremely rare, implying that for most
scintillating sources, longer-term, 
dedicated monitoring programs are required to extract detailed information on 
source structures. 
}
   \maketitle
%
%________________________________________________________________

\section{Introduction}
Since the discovery of inter-day and intraday flux density variations
(IDV) in extragalactic radio sources \citep{hee84,wit86,hee87}, observations
have shown that a significant fraction of all flat-spectrum radio
sources exhibit such variations at frequencies of several GHz
\citep{qui92,ked2001,lov2003}. For a number of sources, it has been
possible to show unambiguously that the rapid variations are due
to interstellar scintillation (ISS) and to constrain scintillation
parameters, mainly through two types of
observation: (i) annual modulations in the 
characteristic timescale of variability, and (ii) for the three fastest
scintillators, discussed in Section~\ref{sec-ihv}, through
observations of delays between the variability pattern arrival
times at widely separated telescopes.
Moreover, sources which show intrinsic changes on short timescales are
likely to contain compact components with angular sizes of order 10
microarcseconds ($\mu$as) or less, which will be affected by ISS at cm
wavelengths.  On one hand, ISS may complicate interpretation of radio
variability,  but on the other, the statistics of ISS fluctuations can
be used as an ultra-high resolution probe of source structure \citep{mj2002}. 
%______________________________________________________________

\section{The intra-hour variables \label{sec-ihv}}
Sources which show very rapid IDV allow a useful sample of the
variability pattern to be obtained in a relatively short observing
time. The discoveries of large-amplitude, intra-hour variations in
three sources have thus made important contributions to our understanding
of ISS of extragalactic sources.

PKS 0405$-$385, a quasar at redshift z=1.285,
was the most extremely variable source discovered in the ATCA IDV
Survey of \cite{ked2001}.
\cite{ked97} found that the frequency dependence
of the observed IDV was consistent with what is expected
from ISS, with the break between strong and weak scattering occurring
near 5~GHz. 
Unequivocal evidence that the rapid variations in
PKS~0405$-$385 are a result of ISS came from the 
detection of a time delay between the variability pattern arrival times
at the VLA and the ATCA \citep{vsop:jau2000}. Such time delay measurements are
only possible when changes in flux density can be detected on
timescales of order tens of seconds.  Subsequently, a detailed
analysis of the scintillation of Stokes I, Q, and U parameters was used
to construct a model of the $\mu$as-scale polarized structure of this
source \citep{ric2002}.   This analysis also found that scattering in
an unusually nearby screen, only 3--30~pc from the observer, is
required to explain the observations. A source brightness
temperature, $T_b$, of $\sim 2\times 10^{13}$~K is then implied. 
The rapid ISS of PKS 0405$-$385 is episodic, lasting
for several weeks to months at a time.
Between these episodes, the source shows large variations on
timescales of months to years, but appears to be stable on
short timescales, with observed rms variations not
exceeding 1\% in 48 hours at 4.8 and 8.6\,GHz.  After a quiescent
period of 4 years, PKS~0405$-$385 has recently started scintillating again
(IAUC 8403). 

Even more rapid variations were discovered in the z=0.54 quasar
J1819+3845 by \citet{dtdb2000}. 
Frequent monitoring with the WSRT has shown a persistent annual cycle
in the characteristic timescale of IDV, and  
time delays between the IDV pattern arrival times at the VLA and WSRT
were observed in January 2001 \citep{dtdb2002}. 
\citet{dtdb2003} used the annual modulation in characteristic timescale 
to constrain the peculiar velocity and distance of the scattering
screen,  and to estimate the screen distance (1--12 pc), source size
and brightness temperature. The scintillation pattern
is highly anisotropic, and the favoured source model implies $T_b \sim
10^{12} K$. 

PKS~1257$-$326, at z=1.256, was the third quasar
discovered to show intra-hour variations. Similarly to
J1819+3845, this source has shown persistent rapid 
variations since its discovery in 2000. 
\cite{big2003} presented
results of ATCA monitoring of PKS~1257$-$326 over more than a year,
showing an annual cycle in the characteristic timescale of
variability. The combined results of these and more recent observations
of this source are presented below, and ongoing analysis is discussed.

\subsection{PKS 1257$-$326: ISS geometry, 
  long-term changes and $\mu$as-scale polarized structure}
\begin{figure}
   \centering
   \includegraphics[width=7cm]{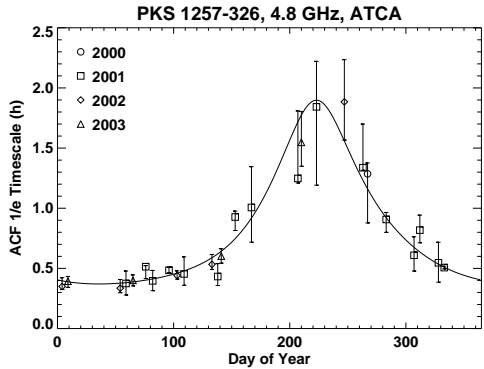}
      \caption{The annual cycle in characteristic timescale 
              for PKS~1257$-$326 at 4.8 GHz. The line shows a
              best fit for the scintillation velocity and an
              elliptical scintillation pattern, using these data 
              combined with two-station time delay data. 
              }
         \label{fig-1257tc}
   \end{figure}

Frequent monitoring of PKS~1257$-$326 with the ATCA was
undertaken from early 2001 until September 2003, revealing a persistent
annual cycle in the characteristic timescale of
IDV. Fig.~\ref{fig-1257tc} shows the characteristic timescale,
defined here as the half-width at $1/e$ of the intensity 
autocorrelation function (ACF), from each ATCA dataset at 4.8~GHz,
plotted against day of year. The methods of \citet{ric2002} was used to
compute estimation errors in the ACFs. Errors are dominated by the limited
sampling of the scintillation pattern. 
%The errors, generally dominated by limited
%sampling of the scintillation pattern, have been estimated using the
%method of \citet{ric2002} to compute estimation errors in the ACFs.

   \begin{figure}
   \centering
   \includegraphics[width=8.8cm]{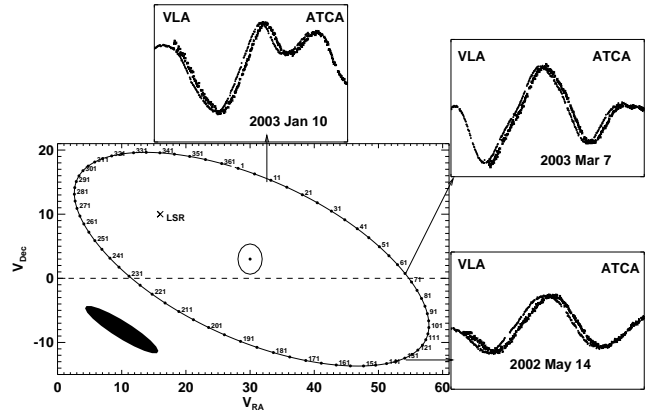}
      \caption{The 2-D scintillation velocity, in km\,s$^{-1}$, 
               over the course of the year for PKS~1257$-$326, 
              and two-station time delay data at 4.9 GHz observed at three
              epochs. The velocity
              ellipse is centered on the best fit screen velocity, with
              error ellipse indicated. The local standard of rest (LSR)
              velocity is also indicated. The small filled ellipse
              shows the orientation of the best fit anisotropic
              scintillation pattern. 
              }
         \label{fig-1257td}
   \end{figure}

Pattern arrival time delays between the VLA and ATCA have been observed for
PKS~1257$-$326 at three different times of the year, on two consecutive
days in each epoch and simultaneously at frequencies of 4.9 and
8.5~GHz. 
We used observations of the nearby
calibration source PKS~1255$-$316 for relative
amplitude calibration of the two arrays. 
PKS~1257$-$326 has faint, arcsecond-scale extended structure, although VLBI
observations show that the source is dominated by an unresolved
component, with no other significant structure visible on mas
scales. In order to compare accurately the flux densities of the
unresolved, scintillating component measured at each array, 
a model of the arcsecond-scale
extended structure, imaged using VLA data, was first subtracted
from the visibilities of both telescopes. 
The light curves observed at both telescopes are essentially identical apart
from a displacement in time. Fig.~\ref{fig-1257td} shows three of the
simultaneous observations at 4.9\,GHz.
The measured time delays were found to be very similar for both
frequencies and on consecutive days. We detect no significant
change in delay over the course of each observation, although a small
change might be expected due to the rotation through $\sim 20^{\circ}$
of the projected ATCA-VLA baseline. For the May 2002 epoch, the best 
fit delay is $470 \pm 60$\,s, while for January and March 2003, the
time delay is shorter, $330 \pm 50$\,s during January and
$320 \pm 40$\,s in March.

The time delay and
annual cycle measurements are combined to fit simultaneously for the
peculiar velocity of the scattering screen, assumed to be the same for 
both frequencies, and the length scale $s_0$, axial ratio $R$, and
position angle of an elliptical scintillation pattern, which was allowed to be
different at the two observed frequencies. Fig.~\ref{fig-1257td}
shows how the scintillation velocity changes over the course of the
year. The large ellipse is centred on the best fit screen velocity.  
The scintillation pattern is found to
be highly anisotropic. The orientation and anisotropy of the fitted 
scintillation pattern is indicated by the small filled ellipse in
Fig.~\ref{fig-1257td}.
The line plotted in Fig.~\ref{fig-1257tc} shows
the annual cycle in characteristic timescale expected from the best fit
model. 
We find $s_0\approx 10^5$\,km at 4.8\,GHz, and $R \geq 5$. 
In weak scattering, $s_0$ corresponds approximately to the size of the
first Fresnel zone, $r_{\rm F}=\sqrt{\lambda L/(2\pi)}$, at the
scattering screen distance $L$.  
Our model then 
implies $L \sim 30$~pc and an angular size of the first Fresnel zone,
$\theta_{\rm F} \sim 20\mu$as. Assuming the intrinsic source
angular size cannot be much larger than this, a brightness temperature
of $\sim 10^{13}$~K is implied. Further investigation, for example 
analysis of power spectra of the fluctuations, and analysis of
multi-frequency and multi-Stokes data, is needed to determine whether
the high degree of anisotropy in the scintillation pattern is due to an
anisotropic source, or to anisotropic scattering. 
Highly anisotropic scintillation patterns have been found for
all three fast scintillators. \citet{ric2002} argued that the anisotropy
is a property of the scattering screen for the case of 
PKS~0405$-$385, based on the observed power spectrum of the
fluctuations. It is interesting to note that an 
interpretation of parabolic arcs observed in the dynamic secondary spectra of
some pulsars \citep{sti2001}
requires highly anisotropic scattering from compact, isolated clumps of
scattering material in the ISM \citep{wal2004}. 

\begin{figure}
   \centering
   \includegraphics[width=8.8cm]{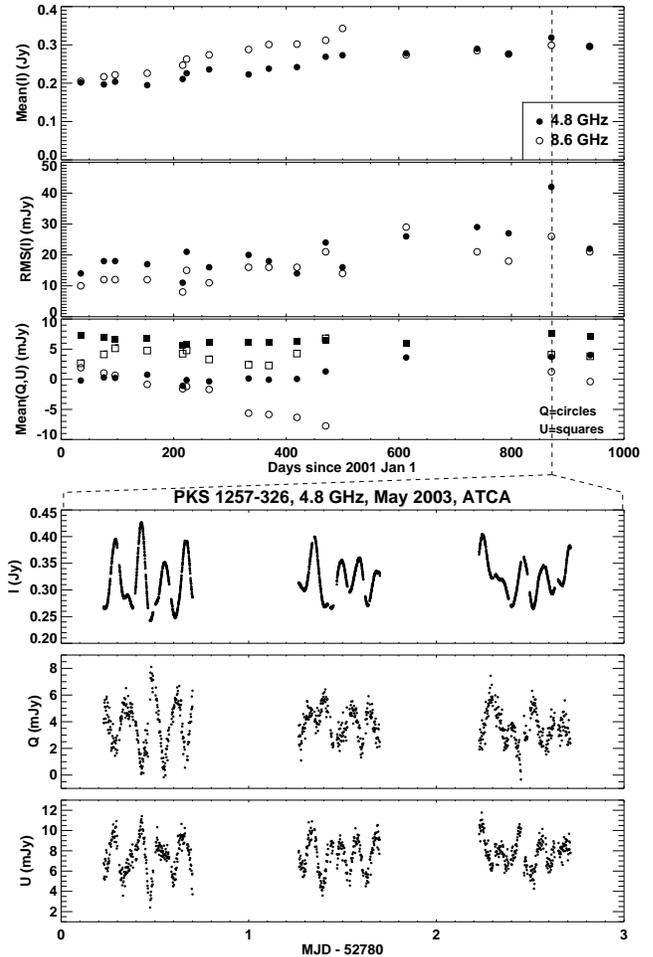}
      \caption{ATCA data for PKS 1257$-$326, showing the average value
   of Stokes I, Q, and U in each observing session at 4.8 and 8.6~GHz
   as well as the rms variations in Stokes I (top, centre panel). The
   duration of a typical observing session is 12--48\,h. The
   lower 3 panels show all data for the observation of 21--23 May
   2003, with Stokes I data plotted with 30\,s averaging,
   and Stokes Q and U plotted with 2 minute averaging.
              }
         \label{fig-1257lt}
   \end{figure}

Although there has been no detectable change in the
annual cycle over the three years of monitoring (see
Fig.~\ref{fig-1257tc}), there has been a significant change in the mean
flux density of PKS~1257$-$326. Fig.~\ref{fig-1257lt} shows the
mean flux density (Stokes I) and rms variations in Stokes I observed
in each epoch, as well as the mean flux density in Stokes Q and U for
each epoch, at 4.8 and 8.6~GHz. The data show that the
source has undergone a slow outburst, of the kind typical of
flat-spectrum quasars, peaking first at the higher frequency and later
at the lower frequency, with a corresponding change in
polarization. The scintillating component has become brighter as
indicated by the increase in rms variations, but evidently the
component has not expanded enough to affect the measured 
scintillation timescale. 

Fig.~\ref{fig-1257lt} also shows all data at 4.8~GHz from 21--23 May
2003. Rapid variations are observed in Stokes Q and U, with a
similar timescale to those observed in Stokes I,
suggesting that the polarized sub-structure of PKS~1257$-$326 is
relatively simple. Preliminary analysis of cross-correlations between the
polarized and total intensity scintillation patterns
indicates that the peak in polarized flux density is displaced from the
peak in total intensity, at both 4.8 and 8.6 GHz, by a distance 
of order 10~$\mu$as. We have observations over a range of  
different directions of the scintillation velocity (see
Fig.~\ref{fig-1257td}), analagous to cuts at different angles
through the $(u,v)$ plane in synthesis imaging. Cross-correlation
and cross-power spectrum analysis of all available Stokes I, Q, and U
data, and a modelling procedure similar to that used by
\citet{ric2002}, would constrain the $\mu$as-scale polarized structure
of PKS~1257$-$326 and its evolution during the observed outburst. 

%______________________________________________________________

\section{Findings of the MASIV Survey} 
The Micro-Arcsecond Scintillation-Induced Variability (MASIV) Survey
was in part motivated by the recent serendipitous discoveries of the two
fast scintillators, J1819+3845 and PKS~1257$-$326, which were both
too weak to have been included in previous IDV Surveys. 
MASIV observations were carried out during 2002 and early 2003 with
the VLA at 4.9\,GHz. \citet{lov2003} describe the source selection and
observing strategy, and present first results of the Survey. 

Sources which showed rms variations larger than  2$\sigma$ were
selected, where $\sigma$ is the estimated uncertainty due
to errors proportional to flux density (antenna pointing offsets, etc.) as well
as noise and confusion. In total, 525 unresolved sources were observed
in all four sessions. Of 
these, 146 (28\%) showed significant variability in at least one epoch,
while in any given epoch, 11--15\% of sources were found to be variable. 
For a significant fraction of sources, rapid variability appears to be
episodic. A large range of variability timescales was observed. 
For those sources which show persistent rapid variations,
approximately half show evidence of an annual cycle in 
characteristic timescale, 
although there are large uncertainties in timescale
estimates due to the limited sampling of the observations.
The sky distribution of the variable source fraction is significantly
anisotropic.  A preliminary analysis of
modulation indices based on weak scintillation theory using the
\citet{tc93} model for the interstellar electron density indicates that
the peak brightness temperature of scintillating sources is in the
neighbourhood of $10^{12}$\,K. This brightness temperature limit is
similar to those obtained from VLBI observations even though ISS is not
subject to the same angular resolution limits. 
A comparison of the strong and weak source sub-samples showed that 
the fraction of variable sources was significantly larger in the weak source
population. A comparison of mas-scale VLBI morphologies (Ojha et
al. 2004, in press)\nocite{ojh2004} indicates that scintillating
sources have a higher proportion of their flux density in an unresolved
core component, and have smaller overall angular extent than
non-scintillating sources. No other source in the MASIV Survey showed
variations as rapid or modulation indices as large as observed in
J1819+3845. It seems that for most sources, scintillation occurs in
more distant scattering screens; most lines-of-sight do not intersect
the nearby screens responsible for intra-hour variations. 
%______________________________________________________________

\section{Dedicated monitoring: the COSMIC program}
The MASIV Survey showed that most ISS occurs on timescales of order 
days or more. To get the same information as obtained from monitoring
of intra-hour variables requires observations over hundreds of days, 
which is not feasible with current National Facility
instruments. However, bright
sources which show relatively large-amplitude variations can be
monitored with dedicated single-dish telescopes. An example of such a
program is the ``COntinuous Single-dish Monitoring of Intraday
variables at Ceduna'' (COSMIC) program, using the University of
Tasmania's 30\,m antenna at Ceduna. When not participating in VLBI
observations, the telescope is operated remotely to observe a small
sample of strong southern IDV sources at 6.7\,GHz. 
Based on observations of calibration sources,
rms intensity variations larger than $\sim 2$\% are detectable in
sources $\sim 1$~Jy or brighter. 

\begin{figure}
   \centering
   \includegraphics[width=8.8cm]{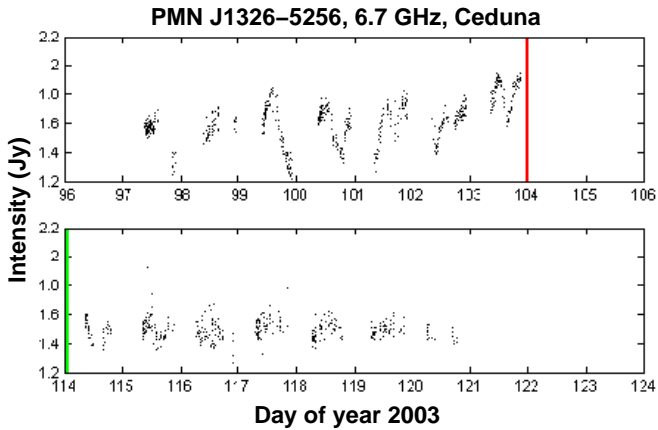}
      \caption{Ceduna data at 6.7~GHz for PMN~J1326$-$5256 data at
   6.7~GHz, showing a change in IDV behaviour over several weeks in
              2003 July. 
              }
         \label{fig-1326ceduna}
   \end{figure}

Analysis and interpretation of the first year of Ceduna data is
ongoing. Various behaviours are observed in the target sources. 
For example, PKS~1519$-$273 shows a persistent pattern of variations with a
repeated annual cycle.
PMN~J1326$-$5253, on the other hand, has shown 
dramatic changes in its IDV properties. After showing large amplitude
IDV for some months, in April 2003 the amplitude of IDV suddenly
decreased, as shown in Fig~\ref{fig-1326ceduna}.  
Further investigation is needed to determine whether observed changes
in scintillation behaviour are due to source-intrinsic or
scattering screen changes.

%______________________________________________________________
\section{Summary and outlook}
ISS of extragalactic sources at cm wavelengths can be used to constrain 
source structure on microarcsecond scales, and potentially allows
measurements of brightness temperatures higher than the $\sim 10^{12}$ K limit
of ground-based VLBI. Dedicated monitoring of small source samples,
required in order to extract detailed information on source structure, are
complementary to large statistical studies such as the MASIV Survey. 
A significant fraction of scintillating sources have relatively stable,
or slowly evolving, long-lived scintillating components, allowing
continued monitoring and modelling of their $\mu$as-scale structure. 
Other sources show dramatic changes with time in their scintillation
behaviour, which may be a result of either intrinsic expansion of a
source component to quench the scintillation, or a change in the effective
scattering along the line-of-sight to the source. 
For a proper interpretation of ISS it is important to consider the
effects of anisotropic scattering, extended source structure, and
scintillation behaviour near the transition  between weak and strong
scattering.  ISS studies, combined with other available data, offer a
unique probe of the physics of inner jets of radio-loud AGN.

\begin{acknowledgements}
The ATCA is part of the Australia Telescope, which is funded by the
Commonwealth of Australia for operation as a National Facility managed
by CSIRO. The VLA is operated by the National Radio Astronomy
Observatory, a facility of the National Science Foundation operated
under cooperative agreement by Associated Universities, Inc. 
\end{acknowledgements}

%\begin{thebibliography}{}

\bibliographystyle{aa}
\bibliography{all}
%
%   \bibitem[1997]{zheng} Zheng, W., Davidsen, A. F., Tytler, D. \& Kriss, G. A.
%      1997, preprint
%\end{thebibliography}

\end{document}